\documentclass[twocolumn]{aastex701}
\usepackage{my_commands}
\usepackage{xcolor}
\usepackage{float}
\usepackage{amsmath}

\begin{document}

\title{Accurate  Galaxy Cluster Shear and Mass Calibration for LSST with AnaCal}

\author[orcid=0000-0002-2897-6326,sname='Zhou']{Conghao Zhou}
\affiliation{Department of Physics, University of California, Santa Cruz, CA 95064, USA}
\email[show]{zhou.conghao@ucsc.edu}

\author[orcid=0000-0003-2880-5102,gname=Xiangchong, sname='Li']{Xiangchong Li}
\affiliation{Brookhaven National Laboratory, Bldg 510, Upton, New York 11973, USA}
\email{xli6@bnl.gov}

\author[orcid=0000-0002-7904-1707,gname=Hao-Yi, sname='Wu']{Hao-Yi Wu}
\affiliation{Department of Physics, Southern Methodist University, Dallas, Texas 75205, USA
}
\email{}

\author[orcid=0000-0002-3881-7724,gname=Anja, sname='von der Linden']{Anja von der Linden}
\affiliation{Department of Physics and Astronomy, Stony Brook University, Stony Brook, NY, 11794, USA}
\email{}

\author[orcid=0000-0001-6089-0365,gname=Jeltema, sname='Tesla']{Tesla Jeltema}
\affiliation{Department of Physics, University of California, Santa Cruz, CA 95064, USA}
\email{}

\author[orcid=0000-0002-6389-5409,gname=Tae-hyeon, sname='Shin']{Tae-hyeon Shin}
\affiliation{Department of Physics and Astronomy, Carnegie Mellon University,
5000 Forbes Ave, Pittsburgh, PA 15213, USA}
\email{}

\author[orcid=0000-0003-3195-5507,gname=Birrer, sname='Simon']{Simon Birrer}
\affiliation{Department of Physics and Astronomy, Stony Brook University, Stony Brook, NY, 11794, USA}
\email{}

\author[orcid=0009-0004-6387-5784,gname=Tomomi, sname='Sunayama']{Tomomi Sunayama}
\affiliation{Academia Sinica Institute of Astronomy and Astrophysics (ASIAA), No.1, Sec. 4, Roosevelt Rd, Taipei 106319, Taiwan, R.O.C}
\email{}

\author[orcid=0000-0001-5422-1958,gname=Shenming, sname='Fu']{Shenming Fu}
\affiliation{NSF-DOE Vera C. Rubin Observatory / SLAC National Accelerator Laboratory, 2575 Sand Hill Road, Menlo Park, CA 94025, USA}
\affiliation{Kavli Institute for Particle Astrophysics \& Cosmology, P. O. Box 2450, Stanford University, Stanford, CA 94305, USA}
\email{}

\author[orcid=0000-0001-9431-3806,gname=Prakruth, sname='Adari']{Prakruth Adari}
\affiliation{Department of Physics and Astronomy, Stony Brook University, Stony Brook, NY, 11794, USA}
\email{}

\author[orcid=0000-0002-1518-0150,gname=Lucie, sname='Baumont']{Lucie Baumont}
\affiliation{Dipartimento di Fisica - Sezione di Astronomia, Università di Trieste, Via Tiepolo 11, 34131 Trieste, Italy}
\affiliation{INAF-Osservatorio Astronomico di Trieste, Via G. B. Tiepolo 11, 34143 Trieste, Italy}
\email{}

\author[orcid=0000-0002-2986-2371,gname=Surhud, sname='More']{Surhud More}
\affiliation{Inter-University Centre for Astronomy and Astrophysics, Post Bag 4, Ganeshkhind,
Pune 411 007, India}
\email{}

\author[orcid=0000-0003-2314-5336,gname=Anthony, sname='Englert']{Anthony Englert}
\affiliation{Department of Physics, Brown University, Providence, RI 02912, USA}
\email{}

\author[orcid=0000-0002-3135-3824,gname=Miranda, sname='Gorsuch']{Miranda Gorsuch}
\affiliation{Physics Department, 2320 Chamberlin Hall, University of Wisconsin-Madison, 1150 University Avenue, Madison, WI 53706-1390, USA}
\email{}

\author[orcid=0000-0002-2598-0514,gname=Andrés, sname='Malagón']{Andrés A. Plazas Malagón}
\affiliation{NSF-DOE Vera C. Rubin Observatory / SLAC National Accelerator Laboratory, 2575 Sand Hill Road, Menlo Park, CA 94025, USA}
\email{}

\collaboration{all}{The LSST DESC Collaboration}

\begin{abstract}
The observed abundance of galaxy clusters as a function of mass and redshift provides a powerful route to precision cosmology; a key challenge for cluster cosmology is to establish the relation between cluster observables and cluster masses, for which cluster weak gravitational lensing has become the standard tool. A key challenge for cluster lensing is that the shear signal near cluster centers can reach the non-linear regime, where many shear estimators rely on perturbative assumptions that must be explicitly validated. In this work, we use image simulations to test the performance of the shear estimator \anacal for cluster weak lensing under conditions representative of the 10-year LSST data. We find that \anacal recovers the input shear with minimal bias even at mildly high shear, $|g|\sim 0.15$. We discover a radially decreasing mean shear response as seen previously in data, driven by the radial dependence of the convergence field; if unmodeled, this effect can bias shear inference. We also find a positive shear-estimation bias at third order in the reduced shear near the cluster center. However, because only a small fraction of galaxies lie in the high-shear regime and those measurements are further downweighted by the covariance matrix, the resulting mean cluster-mass bias for cluster lens masses in $[10^{14} \Msun, 10^{15} \Msun]$---adopting a scale cut of $\sim 0.2$ Mpc at $z=0.25$ ---is 
$0.24 \pm 0.26\%$ under ideal settings.  These results demonstrate that \anacal is a robust tool for accurate cluster mass calibration in the LSST era.
\end{abstract}

\section{Introduction}

Galaxy clusters form from the highest peaks of the primordial density field \citep{bardeenStatisticsPeaksGaussian1986, cenGaussianPeaksClusters1998, kravtsovFormationGalaxyClusters2012}. The mass distribution of galaxy clusters can therefore constrain the overall matter density and the clustering of matter in the universe \citep{fanDeterminingAmplitudeMass1997, vikhlininChandraClusterCosmology2009, mantzObservedGrowthMassive2010, rozoCosmologicalConstraintsSloan2010, allenCosmologicalParametersObservations2011, mantzWeighingGiantsIV2015, bocquetClusterCosmologyConstraints2019, abbottDarkEnergySurvey2020, xhakajClusterCosmologyCluster2023, ghirardiniSRGEROSITAAllSky2024, bocquetSPTClustersHST2024b, abbottDarkEnergySurvey2025, bocquetMultiprobeCosmologyAbundance2025, salcedoCosmologicalConstraintsDark2025}. However, the mass of galaxy clusters is not directly observable, and therefore, for cluster cosmology, it is necessary to establish an accurate relation between observables and cluster masses; this process is known as mass calibration of galaxy clusters \citep{reyesImprovedOpticalMass2008, applegateWeighingGiantsIII2014, vonderlindenRobustWeaklensingMass2014, geachClusterRichnessMass2017, melchiorWeaklensingMassCalibration2017, simetWeakLensingMeasurement2017, mcclintockDarkEnergySurvey2019, ansarinejadMassCalibrationYear32024, fuLoVoCCSIIWeak2024}. Weak gravitational lensing, the coherent small distortion of background source images by the foreground lens  \citep{schneiderGravitationalLenses1992, bartelmannWeakGravitationalLensing2001, mandelbaumWeakLensingPrecision2018}, is an excellent method for cluster mass calibration \citep{johnstonCrosscorrelationWeakLensing2007, mandelbaumPrecisionClusterMass2010, beckerACCURACYWEAKLENSINGCLUSTER2011, rozoStackedWeakLensing2011, hoekstraMassesGalaxyClusters2013, melchiorWeaklensingMassCalibration2017, simetWeakLensingMeasurement2017, murataConstraintsMassRichnessRelation2018, bellagambaAMICOGalaxyClusters2019, mcclintockDarkEnergySurvey2019, miyatakeWeaklensingMassCalibration2019, umetsuClustergalaxyWeakLensing2020, aguenaCLMMLSSTDESCCluster2021, chiuEROSITAFinalEquatorialDepth2022, grandisSRGEROSITAAllSky2024, chiuSRGEROSITAAllSky2025, kleinebreilSRGEROSITAAllSky2025, shinWeakLensingMass2025}, mainly for three reasons. First, gravitational lensing is sensitive to all masses in galaxy clusters. Therefore, it can probe the mass of the dark matter halo without  the need for calibrating the relation between baryonic mass and total mass \citep{bartelmannDarkUniverse2010,vonderlindenRobustWeaklensingMass2014}. Second, weak lensing does not depend on the dynamical or thermal equilibrium assumption of the intracluster medium, unlike the mass estimates based on the X-ray or thermal Sunyaev-Zeldovich effect (SZe), and therefore provides a direct mass calibration \citep{vonderlindenRobustWeaklensingMass2014, mantzWeighingGiantsGalaxy2016, applegateCosmologyAstrophysicsRelaxed2016}. Third, stacked cluster weak gravitational lensing can yield high signal-to-noise measurements of lensing signals around galaxy clusters and enable precise estimations of the mean mass of galaxy cluster populations \citep{oguriCombiningClusterObservables2011, rozoStackedWeakLensing2011, wuCosmologyGalaxyCluster2021}. Because of these advantages, galaxy clusters identified in sub-millimeter, optical/near-infrared, and X-ray surveys routinely use weak gravitational lensing to calibrate their mass-observable relations and, in turn, to constrain cosmology \citep{leauthaudWeakLensingStudy2010, applegateWeighingGiantsIII2014, okabeLoCuSSWeaklensingMass2016, miyatakeWeaklensingMassCalibration2019, costanziCosmologicalConstraintsY12021, schrabbackMassCalibrationDistant2021, chiuEROSITAFinalEquatorialDepth2022, bocquetSPTClustersHST2024a, bocquetSPTClustersHST2024b, ghirardiniSRGEROSITAAllSky2024, grandisSRGEROSITAAllSky2024, robertsonACTDR5SunyaevZeldovichClusters2024, bocquetMultiprobeCosmologyAbundance2025}.

Accurate extraction of weak lensing signals requires inferring the underlying gravitational shear from noisy, PSF-convolved images of lensed source galaxies; this inference process is called shear calibration or shear estimation. The dominant modern paradigm for shear calibration is self-calibration, where the ensemble shear response is estimated directly on the survey images by applying a known shear (or an equivalent perturbation) and measuring the change in the estimator \citep{huffMetacalibrationDirectSelfCalibration2017, sheldonPracticalWeaklensingShear2017}. Examples of this approach are \mcal \citep{huffMetacalibrationDirectSelfCalibration2017, sheldonPracticalWeaklensingShear2017} and its detection-aware extension \mdet \citep{sheldonMitigatingSheardependentObject2020, sheldonMetadetectionWeakLensing2023, yamamotoDarkEnergySurvey2025}.

Another shear calibration approach is \anacal{}, which analytically derives the shear response of the full measurement pipeline---including the observables, detection, and selection steps---by differentiating them with respect to controlled shear perturbations using auto-differentiation and a shapelet basis \citep{liWeakGravitationalLensing2022, liDifferentiablePerturbationbasedWeak2024, liAnalyticalWeakLensing2024}. \anacal and \mdet are selected by the Legacy Survey of Space and Time (LSST) \citep{ivezicLSSTScienceDrivers2019} Dark Energy Science Collaboration (DESC) of the Vera C. Rubin Observatory as the fiducial shear estimation methods: both maintain multiplicative bias $|m|<0.3\%$ in the typical weak-lensing regime for blended galaxies within the same narrow redshift slice \citep{sheldonMetadetectionWeakLensing2023, liAnalyticalWeakLensing2024}, satisfying the requirements in the DESC Science Requirements Document (SRD) \citep{thelsstdarkenergysciencecollaborationLSSTDarkEnergy2018}.

In the standard shear estimation tests \citep{sheldonMetadetectionWeakLensing2023, liAnalyticalWeakLensing2024}, when evaluating the performance of shear estimations, usually only isolated galaxy simulations with a spatial-invariant constant shear that typically reflects the magnitude of large-scale cosmic shear measurements were applied. These fiducial simulations are not optimized for testing shear estimation for cluster lenses for the following reasons. First, cluster lensing probes a high-shear regime that is not well covered by typical weak-lensing validation tests: for an intermediate-mass cluster relevant to cosmology samples, the tangential shear can reach $0.15$ in the inner region around 0.2 Mpc, several times larger than the small-shear values commonly used in perturbation-based calibration tests \citep{okabeLoCuSSWeaklensingMass2016, umetsuClustergalaxyWeakLensing2020, liWeakLensingConstraints2025}. This motivates explicitly validating perturbation-based shear estimation and calibration at high shear. Second, because the cluster lensing deflection field shifts the apparent positions of background galaxies—altering their overlap and blending configuration, which ultimately impacts both detection and shape measurements—we need to test how this inhomogeneous deflection field propagates into shear-estimation bias. Third, near the brightest central galaxy of a galaxy cluster, there is diffuse, low surface brightness stellar halo and intra-cluster light that aligns with the shape of the central galaxy \citep{gonzalezIntraclusterLightNearby2005, zibettiIntergalacticStars0252005, huangIndividualStellarHaloes2018, wangStellarHaloIsolated2019, zhangDarkEnergySurvey2019, broughVeraRubinObservatory2020, klugePhotometricDissectionIntracluster2021, liReachingEdgeProbing2022, montesFaintLightGroups2022, zhangDarkEnergySurvey2024, englertIntraclusterLightAbell2025}. If this anisotropic light is not taken into account in the shape measurement, it might introduce a  spurious alignment signal and attenuate the lensing signal \citep{zhouIntrinsicAlignmentRed2023}. Fourth, foreground cluster members may blend with background source galaxies and bias shape measurement and photometric redshift calibration \citep{vargaDarkEnergySurvey2019, maccrannDarkEnergySurvey2021}. This effect is strongest near cluster center since the cluster number density radially decreases from the center. Finally, the distortion of gravitational lensing is not limited to shear and convergence. There are higher-order distortions, such as flexion, that might not be negligible when the gravitational effect is high \citep{baconWeakGravitationalFlexion2006, hawkenGravitationalFlexionElliptical2009, violaShearflexionCrosstalkWeaklensing2012,liuMeasurementCalibrationNonlinear2024}.

All the above effects may bias shear estimation for cluster lenses, and these sources of error can couple non-trivially. Systematically isolating and testing them is therefore essential for understanding their individual contributions. In this paper, we focus on testing \anacal's robustness against high shear and inhomogeneous deflection field, and therefore we do not include effects caused by higher-order distortion (e.g., flexion \citep{baconWeakGravitationalFlexion2006}), cluster member galaxies, intra-cluster light, and redshift calibration in the simulations discussed in this paper. The paper is organized as follows: In Section~\ref{sec:methods}, we introduce \anacal shear estimation framework, the cluster lensing simulation, and various methods to evaluate the estimated shear. In Section~\ref{sec:result}, we present our findings, such as shear bias, mass bias, and shear response trends. In Section~\ref{sec:summary}, we compare our results with previous literature, summarize our results, and discuss potential caveats and future plans. We use a flat $\Lambda$CDM cosmology from \citet{collaborationPlanck2018Results2020}, adopting $\Omega_{\rm m} = 0.315$, $\Omega_{\Lambda} = 0.685$, and $H_0 = 67.4\ \mathrm{km\ s^{-1}\ Mpc^{-1}}$.

\section{Methods}
\label{sec:methods}
\subsection{AnaCal shear estimator}
In this section, we briefly describe shear estimation with \fpfs shapes
\citep{liFourierPowerFunction2018,liWeakGravitationalLensing2022} calibrated
with \anacal \citep{liAnalyticalWeaklensingShear2023,
liAnalyticalWeakLensing2024,liAnalyticalWeaklensingShear2025}. Following
\citet{liAnalyticalWeakLensing2024}, we define $\overline{\boldsymbol{\nu}},
\boldsymbol{\nu}$ and $\tilde{\boldsymbol{\nu}}$ as the prelensed, lensed
noiseless and lensed noisy linear observables of the galaxy. In \fpfs, images
are projected onto a set of basis kernels to form linear observable modes.
These basis kernels are composed of detection kernels and shapelet kernels \citep{masseyPolarShapelets2005},
which jointly capture the information used for object detection and for
estimating galaxy shape, size, and flux. Mathematically, this projection can be
written as:
\begin{equation}
    \nu_i=\int_{\boldsymbol{k}} f(\boldsymbol{k}) \frac{\chi_i(\boldsymbol{k})}{p(\boldsymbol{k})},
\end{equation}
where $\chi_i(\boldsymbol{k})$ is the basis kernel, $p(\boldsymbol{k})$ is PSF,
and $f(\boldsymbol{k})$ is the galaxy image in Fourier space.

The peak-detection basis kernels, used for object detection and for computing
the corresponding shear response, are given by
\begin{equation}
    \psi_i=\frac{1}{(2 \pi)^2} e^{-|\boldsymbol{k}|^2 \sigma_h^2 / 2}\left(1-e^{\mathrm{i}\left(k_1 x_i+k_2 y_i\right)}\right),
\end{equation}
where $\left(x_i, y_i\right)=(\cos (i \pi / 2), \sin (i \pi / 2))$, and $i \in \{0,1,2,3\}$.

To measure galaxy properties and their shear response, we use polar shapelets:
\begin{equation}
    \begin{aligned} \phi_{n m}\left(\boldsymbol{x} \mid \sigma_h\right) & =(-1)^{(n-|m|) / 2}\left\{\frac{[(n-|m|) / 2]!}{[(n+|m|) / 2]!}\right\}^{\frac{1}{2}} \\ & \times\left(\frac{\rho}{\sigma_h}\right)^{|m|} L_{\frac{n-|m|}{2}}^{|m|}\left(\frac{\rho^2}{\sigma_h^2}\right) e^{-\rho^2 / 2 \sigma_h^2} e^{i m \theta}\end{aligned},
\end{equation}
where $L_{\frac{n-|m|}{2}}^{|m|}$ are the Laguerre polynomials, $n$ is a non-negative integer radial number, $m$ is the spin number with $m \in \{-n, -n+1, \ldots n-1, n\}$, $\rho$ is the radial coordinate in a 2D polar coordinate system, and $\sigma_h$ is the characteristic scale radius of the shapelet basis functions.

When shear is small, under the first-order approximation, linear observables transform as
\begin{equation}
    \boldsymbol{\nu}
    =\left(\overline{\boldsymbol{\nu}}+\gamma_1 \boldsymbol{\nu}_{; 1}
    +\gamma_2 \boldsymbol{\nu}_{; 2}\right)+\mathcal{O}\left(\gamma^2\right),
\end{equation}
where $\boldsymbol{\nu}_{; 1}$ and $\boldsymbol{\nu}_{; 2}$ are the linear
shear response defined as
\begin{equation}
    \boldsymbol{\nu}_{; 1} \equiv \frac{\partial \boldsymbol{\nu}}{\partial \gamma_1}, \quad \quad \boldsymbol{\nu}_{; 2} \equiv \frac{\partial \boldsymbol{\nu}}{\partial \gamma_2}.
\end{equation}
The linear shear response can be written as the projection of the deconvolved
image onto the shear response of the basis kernel
\begin{equation}
    \nu_{; 1 i}=\int_{\boldsymbol{k}} f(\boldsymbol{k}) \frac{\chi_{; 1 i}(\boldsymbol{k})}{p(\boldsymbol{k})}, \quad \nu_{; 2 i}=\int_{\boldsymbol{k}} f(\boldsymbol{k}) \frac{\chi_{; 2 i}(\boldsymbol{k})}{p(\boldsymbol{k})}.
\end{equation}
For both peak detection basis kernel polar shapelets, the derivatives with
respect to shear have closed forms. Therefore, the shear response of linear
observables can be written analytically in terms of the shear response of the basis
kernels if we know the projection of images onto the basis kernels.

The weighted galaxy ellipticity $e_{1,2}$ are nonlinear functions of the linear
observables $\boldsymbol{\nu}$:
\begin{equation}
    e_{1,2}(\boldsymbol{\nu})=\epsilon_{1,2}(\boldsymbol{\nu}) w_s(\boldsymbol{\nu}) w_d(\boldsymbol{\nu}),
\end{equation}
where $\epsilon_{1,2}$ is the \fpfs ellipticity
\citep{liFourierPowerFunction2018}, $w_s$ is the selection weight, and $w_d$ is
the detection weight. \fpfs ellipticity is defined as
\begin{equation}
    \epsilon_1+\mathrm{i} \epsilon_2 \equiv \frac{M_{22}}{M_{00}+C},
\end{equation}
where $M_{00}$ and $M_{22}$ are the linear observables from projecting the
galaxy image onto polar shapelet basis $\phi_{00}$ and $\phi_{22}$, and $C$ is
a weighting parameter that controls the relative weights assigned to galaxies
of different brightness.

The selection weight $w_s$ can be written as
\begin{equation}
    w_s = w_0w_2,
\end{equation}
where $w_0$ is the selection weight on SNR and $w_2$ is the selection weight on galaxy size. For a selection with minimal SNR $s_{\mathrm{min}}$, $w_0$ is
\begin{equation}
    w_0=\zeta_{\Omega_0}\left(\frac{M_{00}}{\sigma_0}-s_{\min }\right),
\end{equation}
where $\sigma_0$ is the standard deviation of measurement error due to image noise on the zeroth order shapelet mode $M_{00}$. For a selection with minimal galaxy size $r_{\mathrm{min}}$, $w_2$ is
\begin{equation}
    w_2=\zeta_{\Omega_2}\left(M_{20}+\left(1-r_{\min }\right) M_{00}\right),
\end{equation}
where $\zeta_\Omega$ is the smoothstep function with smoothness parameter $\Omega$:
\begin{equation}
    \zeta_\Omega(x) =
    \begin{cases}
        6\!\left(\frac{x+\Omega}{2 \Omega}\right)^5
        \!\!-\!\!15\!\left(\frac{x+\Omega}{2 \Omega}\right)^4
        \!\!+\!\!10\!\left(\frac{x+\Omega}{2 \Omega}\right)^3 &
        x \in [-\Omega, \Omega]\\
        0 \; &\mathrm{else}\,.
    \end{cases}
\end{equation}

The detection weight $w_d$ is
\begin{equation}
    w_d=\zeta_{\Omega_d}\left(\prod_{i=0}^3 \zeta_{\Omega_q}\left(q_i-\Omega_q-0.8 \sigma_q \right)-w_{\min }\right),
\end{equation}
where $q_i$ is the projection of the galaxy image onto the peak detection basis kernel $\psi_i$, and $\sigma_q$ is the standard deviation of the measurement error on the peak detection modes. The default hyperparameters are chosen to maximize effective galaxy number density in image simulations. We summarize the effective hyperparameters in Table~\ref{tab:hyperparams}.

\begin{table}[h]
    \centering
    \begin{tabular}{|c| c|}
        \hline
        Parameter & Value \\
        \hline
        $C$ & $4\sigma_0$ \\
        $\Omega_0 = \Omega_2$ & $1.6\sigma_0$ \\
        $\Omega_q$ & $1.6\sigma_q$ \\
        $\Omega_d$ & $0.04$ \\
        $s_{\text{min}}$ & $12$ \\
        $r_{\text{min}}$ & $0.1$ \\
        $w_{\text{min}}$ & $0.12$ \\
        \hline
        $H_0$ & $67.66 \text{ km s}^{-1} \text{Mpc}^{-1}$ \\
        $\Omega_{m,0}$ & $0.30966$ \\
        $\Omega_{b,0}$ & $0.04897$ \\
        $T_{\text{CMB},0}$ & $2.7255 \text{ K}$ \\
        $N_{\text{eff}}$ & $3.046$ \\
        $m_\nu$ & $[0, 0, 0.06] \text{ eV}$ \\
        \hline
    \end{tabular}
    \caption{Input \anacal hyperparameters and cosmological parameters and their values used in this paper.}
    \label{tab:hyperparams}
\end{table}

Since the selection and detection weights are chosen to be invariant under $90$-degrees rotations and the galaxy ellipticity is spin-2, the weighted ellipticity remains a spin-2 quantity. We assume the source galaxies orient isotropically and are uniformly distributed prior to shear. The mean weighted ellipticity is
\begin{equation}
    \left\langle\left. e_{1,2}\right|_{\gamma=0}\right\rangle_g=0,
\end{equation}
where $\langle\cdot\rangle_g$ is averaging over the galaxy sample. Since the weighted ellipticity is spin-2, the average of second order
\begin{equation}
    \left\langle\left.\frac{\partial^2 e_{1,2}}{\partial \gamma_i \partial \gamma_j}\right|_{\gamma=0}\right\rangle_g=0.
\end{equation}

Therefore, we can get the mean response with
\begin{equation}
\left\langle\frac{\partial e_i}{\partial \gamma_j}\right\rangle=\left\langle\left.\frac{\partial e_i}{\partial \gamma_j}\right|_{\gamma=0}\right\rangle_g+\mathcal{O}\left(\gamma^2\right).
\end{equation}

The relation between the mean ellipticity and the applied shear is

\begin{equation}
    \left\langle e_i\right\rangle=\sum_{j=1}^2\left\langle\frac{\partial e_i}{\partial \gamma_j}\right\rangle_g \gamma_j+\mathcal{O}\left(\gamma^3\right).
\end{equation}

The components of the linear shear response matrix $\mathbf{R}$ can be obtained by applying the chain rule
\begin{equation}
    R_{ij} \equiv \frac{\partial e_i}{\partial \gamma_j}=\sum_k \frac{\partial e_i}{\partial \nu_k} \frac{\partial \nu_k}{\partial \gamma_j}=\sum_k \frac{\partial e_i}{\partial \nu_k} \nu_{; j k}
\end{equation}

\anacal{} employs a ``renoising'' scheme to correct noise bias in shear
estimation. The idea is to add a controlled, synthetic noise realization to
each image. The synthetic noise has the same noise properties as the original
image noise on the pre-lensed plane after a 90-degree rotation
\citep{sheldonPracticalWeaklensingShear2017}.
\citet{liAnalyticalWeakLensing2024} provides a formal analytic treatment within
the \anacal{} framework and computes the shear response directly from the
renoised observables.

\subsection{Image simulations}

We use several open-source software packages---\descwlsims \citep{sheldonMitigatingSheardependentObject2020, sheldonMetadetectionWeakLensing2023} and \xlens\footnote{https://github.com/mr-superonion/xlens}---to generate realistic cluster-lensing image simulations and to measure shear. \descwlsims \citep{sheldonMetadetectionWeakLensing2023} is built on \galsim \citep{roweGALSIMModularGalaxy2015} and produces calibrated, LSST Science Pipelines–like exposures from an input catalog of galaxies and stars, rendering multi-epoch, multi-band images with user-specified WCS, PSF models, noise, image artifacts, applied shear, and object layouts. \xlens provides an LSST-style modular workflow with three stages—simulation, processing, and analysis. The simulation stage uses \descwlsims to generate images under different lensing scenarios (including both constant-shear tests and cluster lensing). The processing stage runs a chosen shear-estimation method on the simulated images, and the analysis stage summarizes the outputs and produces diagnostic plots. This modular structure enables (i) consistent end-to-end comparisons between cosmic-shear and cluster-lensing configurations, (ii) direct performance comparisons among shear estimators by swapping the processing stage while holding the simulations fixed, and (iii) straightforward application of the same processing/analysis stages to calibrated observational data once validated on simulations.

For this work, we simulate cluster-lensing scenes for  9 halo masses spanning from $\mtwoh=10^{14}\,\Msun$ to $10^{15}\,\Msun$, roughly corresponding to optical richness \citep{rykoffRedMaPPerAlgorithmSDSS2014, rykoffRedMaPPerGalaxyCluster2016} $\lambda \sim 15$--$200$. Each cluster is modeled as an NFW halo \citep{navarroUniversalDensityProfile1997}, and for each mass, we assign an NFW concentration using the mass–concentration relation of \citet{ishiyamaUchuuSimulationsData2021}, evaluated with the \textsc{colossus} implementation \citep{diemerCOLOSSUSPythonToolkit2018}; the resulting $(\mtwoh,\ctwoh)$ values are listed in Table~\ref{tab:mass_table}. All clusters are placed at $z_l=0.25$, and sources are in a single plane at $z_s=1.0$, which roughly corresponds to the median redshift of the LSST 10-year source sample \citep{lsstdarkenergysciencecollaborationlsstdescLSSTDESCDC22021}. We generate 2178 realizations per mass, each covering a $5000\times 5000$ pixel field with pixel scale $0\farcs2$ (i.e., $\approx 0.278^\circ \times 0.278^\circ$, $\approx 0.077 \, \mathrm{deg}^2$, or $\approx 4$ Mpc $\times$ 4 Mpc per realization), corresponding to $\approx 1700 \, \mathrm{deg}^2$ per mass and $\approx 1800 \, \mathrm{deg}^2$ in total across all masses (counting one image from each rotated pair). For each realization, we also simulate a paired image in which all galaxies are rotated by $90^\circ$, enabling effective shape-noise cancellation \citep{pujolHighlyPreciseShear2019, sheldonMetadetectionWeakLensing2023}. We adopt a constant circular Moffat PSF with $\beta=2.5$ and ${\rm FWHM}=0\farcs8$, representative of expected Rubin image quality \citep{lsstdarkenergysciencecollaborationlsstdescLSSTDESCDC22021}.

Source galaxies are drawn from the CatSim-based models \citep{connollySimulatingLSSTSystem2010, connollyEndtoendSimulationFramework2014} used in LSST DESC Data Challenge 1 \citep{sanchezLSSTDESCData2020} and calibrated to  Hyper Suprime-Cam data \citep{aiharaSecondDataRelease2019}. The catalog has a raw source density of  $\sim 240~\mathrm{arcmin}^{-2}$. The effective source number density $n_{\rm eff}$ is $\sim 17 \; \mathrm{arcmin}^{-2}$ \citep{liAnalyticalNoiseBias2025, parkDeepfieldAnalyticalCalibration2026}. The models include bulge+disk+AGN components and reach an effective $i$-band AB magnitude limit of $\sim 27$. To isolate shear-estimation effects from blending between cluster members and background sources, we do not include cluster member galaxies in these simulations. We only apply a signal-to-noise cut of 5. We also apply a size cut $M_{20}/M_{00} + 1 > 0.05$ to reduce stellar contamination \citep{liAnalyticalNoiseBias2025}. The resulting catalog has a limiting magnitude of $\sim 25$.

\begin{figure}[h]
    \includegraphics[width=\columnwidth]{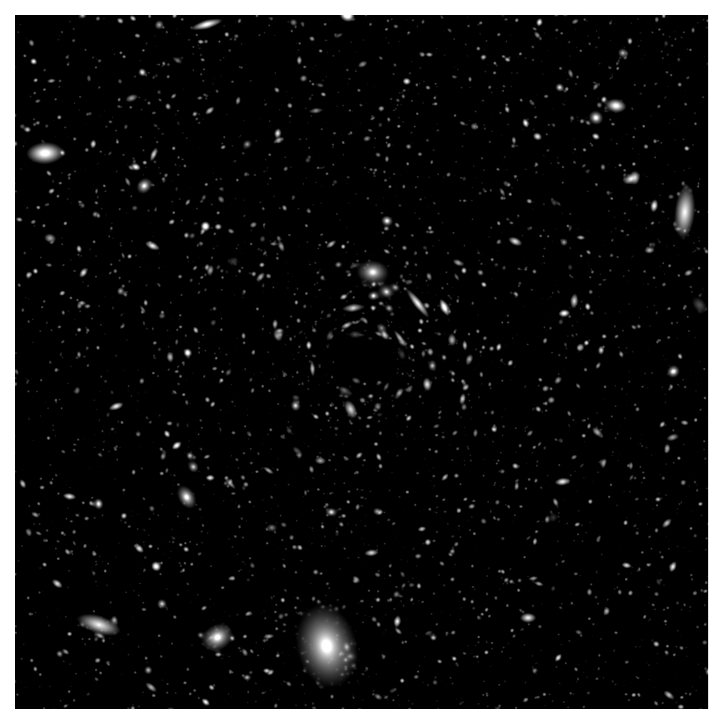}
    \caption{Example simulated image for weak gravitational lensing by a cluster with mass of $10^{15} \Msun$ in LSST. The lens cluster is located at the center of the image, and only source galaxies are simulated. The image size is $5000 \times 5000$ pixels, and the pixel scale is 0.2 arcsec/pixel. We only show the central $2000 \times 2000$ pixels in this image.}
    \label{fig:ex_sim}
\end{figure}

We compute the lensing deflection, shear, and convergence fields using \lenstronomy \citep{birrerLenstronomyMultipurposeGravitational2018, birrerLenstronomyIIGravitational2021} for the NFW lens model, and apply the corresponding (local) affine transformations to the source-galaxy images within \galsim. Figure~\ref{fig:ex_sim} shows an example realization for a $10^{15}\,\Msun$ cluster under the simulation setup described above.

\subsection{Cluster lensing}

For simulated source galaxies with $e_1$ and $e_2$ and position angle $\phi$, we calculate tangential ellipticity and cross ellipticity $e_T$ and $e_X$ with

\begin{equation}
\left[\begin{array}{c}
e_T \\
e_X
\end{array}\right]=-\left[\begin{array}{cc}
\cos 2 \phi & \sin 2 \phi \\
-\sin 2 \phi & \cos 2 \phi
\end{array}\right]\left[\begin{array}{l}
e_1 \\
e_2
\end{array}\right] .
\end{equation}

We also rotate the response matrix into the tangential frame with \citep{mcclintockDarkEnergySurvey2019}
\begin{align}
&\mathbf{R}_{T}
\equiv
\begin{bmatrix}
R_T & R_{TX} \\
R_{XT} & R_X
\end{bmatrix} \\
&=
\begin{bmatrix}
\cos 2\phi & \sin 2\phi \\
-\sin 2\phi & \cos 2\phi
\end{bmatrix}
\begin{bmatrix}
R_{11} & R_{12} \\
R_{21} & R_{22}
\end{bmatrix}
\begin{bmatrix}
\cos 2\phi & \sin 2\phi \\
-\sin 2\phi & \cos 2\phi
\end{bmatrix}^{\!\top}.
\end{align}

We checked that the off-diagonal elements of the response matrix, $R_{TX}$ and $R_{XT}$, have
negligible expectation values, so we focus on the diagonal terms:
\begin{equation}
    \left[\begin{array}{l}
R_T \\
R_X
\end{array}\right] = \left[\begin{array}{ll}
\cos ^2 2 \phi & \sin ^2 2 \phi \\
\sin ^2 2 \phi & \cos ^2 2 \phi
\end{array}\right]\left[\begin{array}{l}
R_{11} \\
R_{22}
\end{array}\right]
\end{equation}

The tangential and cross shear in a radial bin is therefore

\begin{equation}
    \langle \gtobs \rangle_b =  \frac{\left\langle e_T \right\rangle_b}{\left\langle R_T \right\rangle_b},
\end{equation}
where $\langle \cdot \rangle_b$ stands for averaging over all source galaxies in an angular bin. Note, instead of using the average of shear responses in all angular bins, we use the shear response in each angular bin separately. Using the average response in all angular bins would lead to a bias in shear estimation since shear response has a radial trend caused by the convergence field, as described in Section~\ref{sec:response_trend}.

\begin{table}[]
\centering
\begin{tabular}{|c|c|c|c|c|c|c|c|}
\hline
$\mtwoh$ [$\Msun$] & $\ctwoh$ & Mean $\lambda$ & $\lambda$ bin   \\ \hline
$1 \times 10^{14}$     &   4.25 & 11.0 & [10,20]            \\ \hline
$2 \times 10^{14}$     &   4.02 & 25.6 & [20,45]            \\ \hline
$5 \times 10^{14}$     &   3.88 & 78.5 & [70,120]         \\ \hline
$8 \times 10^{14}$     &  3.90 & 139.3 &  [120, 200]            \\ \hline
$8.5 \times 10^{14}$     &  3.92 & 149.9 &  [120, 200]            \\ \hline
$9 \times 10^{14}$     &   3.92 & 160.8 &  [120, 200]         \\ \hline
$9.5 \times 10^{14}$     &  3.94 & 171.8 & [120, 200]            \\ \hline
$1 \times 10^{15}$     &   3.95 & 182.9 &  [120, 200]         \\ \hline
$1.1 \times 10^{15}$     &   3.95 & 205.4 &  [200,300]         \\ \hline
\end{tabular}
\caption{Mass, concentration, and mean richness of cluster lens simulated in this paper. The concentration is calculated with the mass-concentration relation from \citet{ishiyamaUchuuSimulationsData2021}, and the mean richness is calculated with a fiducial mass-richness scaling relation from \citet{bocquetSPTClustersHST2024b}.}
\label{tab:mass_table}
\end{table}

\subsection{Cluster lensing covariance}
\label{sec:covariance}
To conduct cluster mass estimation with a realistic covariance matrix, we derive the analytical cluster lensing covariance matrix based on the forecast quantity of 10-year LSST data with the formalism from \citet{wuCovarianceMatricesGalaxy2019}. We briefly describe the calculation of the lensing covariance matrix and refer the readers to \citet{wuCovarianceMatricesGalaxy2019} for more details.

Assuming both halo number density and matter overdensity follow Gaussian random fields, the covariance of $\gamma_T$ between $\theta_1$ and $\theta_2$ is

\begin{equation}
    \begin{array}{r}
\operatorname{Cov}^{\text {Gauss }}\left[\gamma_T\left(\theta_1\right), \gamma_T\left(\theta_2\right)\right]=\frac{1}{4 \pi f_{\text {sky }}} \int \frac{\ell \mathrm{d} \ell}{2 \pi} \hat{J}_2\left(\ell \theta_1\right) \hat{J}_2\left(\ell \theta_2\right) \times \\
{\left[\left(C_{\ell}^{\mathrm{hh}}+\frac{1}{n_{\mathrm{h}}^{(2 \mathrm{D})}}\right)\left(C_{\ell}^{\kappa \kappa}+\frac{\sigma_\gamma^2}{n_{\mathrm{s}}^{(2 \mathrm{D})}}\right)+\left(C_{\ell}^{\mathrm{h} \kappa}\right)^2\right] ,}
\end{array}
\end{equation}
where $\fsky$ is the fractional sky coverage of the survey, $\sigma_\gamma$ is the scatter for shear,  $n_\mathrm{h}^{\mathrm{(2D)}}$, $n_\mathrm{s}^{\mathrm{(2D)}}$ are the cluster and source surface density, $C_l^{\mathrm{hh}}$, $C_l^{\mathrm{h\kappa}}$, $C_l^{\mathrm{\kappa\kappa}}$ are the halo-halo, halo-convergence, convergence-convergence angular power spectra, respectively. $\hat{J}_2$ is the bin-averaged Bessel function of the first kind, used to account for the finite angular bin size,
\begin{equation}
\hat{J}_2\left(\ell, \theta_{\min}, \theta_{\max }\right)=\frac{1}{\pi\left(\theta_{\max }^2-\theta_{\min}^2\right)} \int_{\theta_{\min}}^{\theta_{\max }} J_2(\ell \theta) 2 \pi \theta \mathrm{~d} \theta,
\end{equation}
in which $\theta_{\min}$ and $\theta_{\max}$ are the lower and upper boundaries of the angular bin, and $J_2$ is the order-2 Bessel function of the first kind. We do not include the non-Gaussian components in the covariance since most of our radial bins are in the large scale regime where the Gaussian approximation works well \citep{wuCovarianceMatricesGalaxy2019}. We use the shear rather than the reduced shear in the covariance, where the correction is second order in the convergence and therefore negligible.

We assume the 10-year LSST area to be 18,000 deg$^{2}$ \citep{ivezicLSSTScienceDrivers2019} and the source density $n_s^{\mathrm{(2D)}}$ to be 27/arcmin$^{2}$ \citep{changEffectiveNumberDensity2013, thelsstdarkenergysciencecollaborationLSSTDarkEnergy2018}. To estimate $n_h^{\mathrm{(2D)}}$, $C_l^{\mathrm{hh}}$, and $C_l^{\mathrm{h\kappa}}$, we forecast cluster number counts and mean halo bias with the following mass-richness relations:
\begin{equation}
    \begin{aligned}
\langle\ln \lambda \mid M_{200c} \rangle= & \ln A_\lambda+B_\lambda \ln \left(\frac{M_{200 \mathrm{c}}}{3 \times 10^{14} h^{-1} \mathrm{M}_{\odot}}\right), \\
\ln \lambda \mid M_{200c} \sim & \mathcal{N}\left(\langle \lambda \mid M_{200c} \rangle, \sigma_\lambda \right),
\end{aligned}
\end{equation}
where $\ln A_{\lambda}$ and $B_{\lambda}$ are the intercept and the slope of the mass-richness relation, and $\sigma_\lambda$ is the scatter of the natural log of the richness at given halo mass.

We adopt the maximum-a-posteriori parameter values from \citet{bocquetMultiprobeCosmologyAbundance2025} and consider a single cluster redshift bin, $0.2<z<1.0$. We combine the mass--richness relation with the halo mass function of \citet{tinkerHaloMassFunction2008} to predict the richness distribution, and compute expected counts in richness bins with edges $\lambda = 10, 20, 45, 70, 120, 200, 300$. The redshift and richness binning follow the LSST Science Requirements Document \citep{lsstdarkenergysciencecollaborationlsstdescLSSTDESCDC22021}. Our deliberately conservative binning targets the high signal-to-noise regime, enabling cleaner tests of shear and mass estimation in the presence of  high lensing signals. For a simulated cluster of mass $m$, we assign it the covariance of the richness bin $[\lambda_{\min},\lambda_{\max}]$ when its expected mean richness falls in that interval (equivalently, when $\langle \ln\lambda \mid m\rangle \in [\ln\lambda_{\min},\ln\lambda_{\max}]$). The cluster masses, concentrations, and mean richness values used in our simulations are summarized in Table~\ref{tab:mass_table}.

\subsection{Mass fitting}

We forward model the $g_T$ profile to fit for mass. For a source galaxy with distance $\theta$ to the lens center with response $R_T$, we calculate individual tangential reduced shear $g_T$ with NFW profiles using \lenstronomy. Then, for each radial bin, we calculate the model vector with
\begin{equation}
    \langle \gtmodel \rangle_b (\mtwoh, \ctwoh) = \frac{\sum_i R_{T,i} g_T(\mtwoh, \ctwoh, \theta_i)}{\sum_i R_{T,i}},
\end{equation}
where $\sum_i$ denotes the sum over all source galaxies within the corresponding angular bin. The parameters $\mtwoh$ and $\ctwoh$ represent the mass and concentration of the model halo, $g_T(\mtwoh, \ctwoh, \theta_i)$ is the model reduced tangential shear evaluated at the angular separation $\theta_i$ of the $i$-th source from the cluster center, and $R_{T,i}$ is the tangential shear response of that source. We perform the model fitting under two configurations: one adopting a fixed mass-concentration relation from \citet{ishiyamaUchuuSimulationsData2021}, and another leaving both mass and concentration as free parameters. We use 15 linearly spaced radial bins spanning from 15 to 495 arcsec, which corresponds to physical scales of approximately 0.6 to 2 Mpc at $z = 0.25$. We emphasize that, compared to evaluating the model at the radial bin centers and using a mean shear response, this full forward-modeling approach naturally avoids biases introduced by the intra-bin radial dependence of both the shear response and the source galaxy number density.

We fit the masses with the affine invariant Markov Chain Monte Carlo ensemble sampler \citep{goodmanEnsembleSamplersAffine2010} implemented in \textsc{emcee} \citep{foreman-mackeyEmceeMCMCHammer2013} with the following log likelihood function
\begin{equation}
    l(\mtwoh) = (\gtmodel - \gtobs)^T\mathbf{C}^{-1} (\gtmodel - \gtobs).
\end{equation}

\begin{figure}[h]
    \includegraphics[width=\columnwidth]{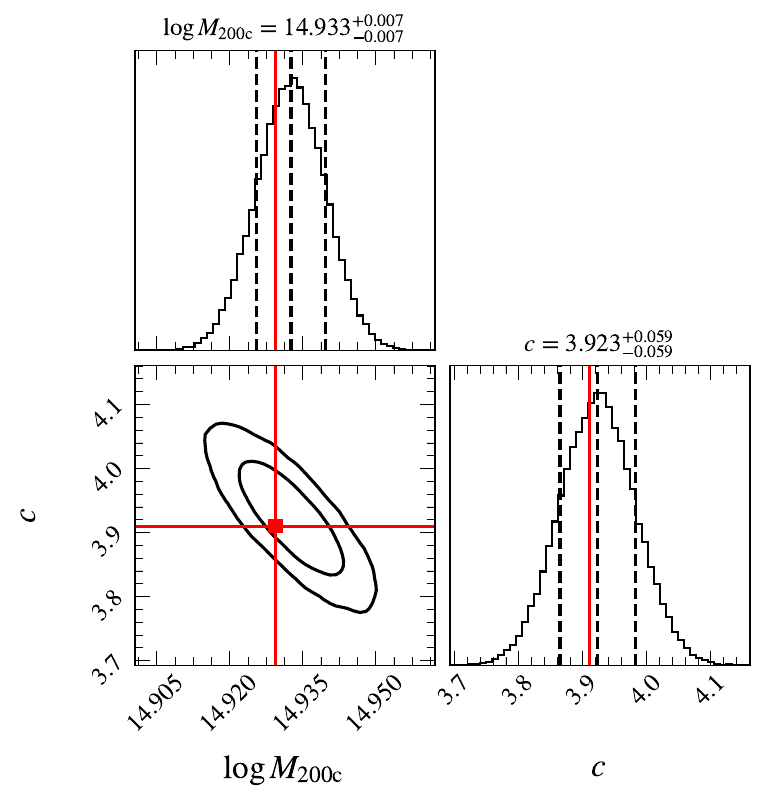}
    \caption{Example posterior distribution for mass fitting for a cluster lens of mass $8.5 \times 10^{14} \Msun$ with free mass concentration relation. The inner and outer contours denote 68\% and 95\% credible regions. }
    \label{fig:ex_corner}
\end{figure}

For each  bin, we run 10000 steps with 32 walkers and burn in 5000 steps. In Figure~\ref{fig:ex_corner}, we show an example posterior distribution for a cluster lens with $\mtwoh = 8.5 \times 10^{14} \Msun$.

\subsection{Shear estimation bias}
\label{sec:shear_bias}
To estimate the shear bias, we first calculate the true shear profile with \lenstronomy with the true input mass, concentration, and source positions.

Then we calculate
\begin{equation}
\label{eq:shear_bias}
    m = \frac{\gtobs}{\gttrue} - 1 = a + b (\gttrue)^2,
\end{equation}
where $m$ is the overall shear multiplicative bias, $a$ is the linear
multiplicative-bias coefficient and $b$ is the leading cubic nonlinearity
(third-order bias) coefficient. The additive bias term is 0, so we do not include it in the fitting. We do not include the second-order bias term
because it carries a different spin from the shear; by symmetry, it cannot
contribute to the shear estimate. In addition, we do not include the $\kappa$
terms since we tested that the shear estimation bias is independent of $\kappa$
as shown in Appendix~\ref{appendix:shear_bias}. We gather the results from all
clusters and fit for $a$ and $b$ with the least square fitting implemented in
\textsc{scipy.optimize.curvefit}.

\section{Results}
\label{sec:result}

\subsection{Accurate shear recovery}

\begin{figure*}[htbp]
    \includegraphics[width=\textwidth]{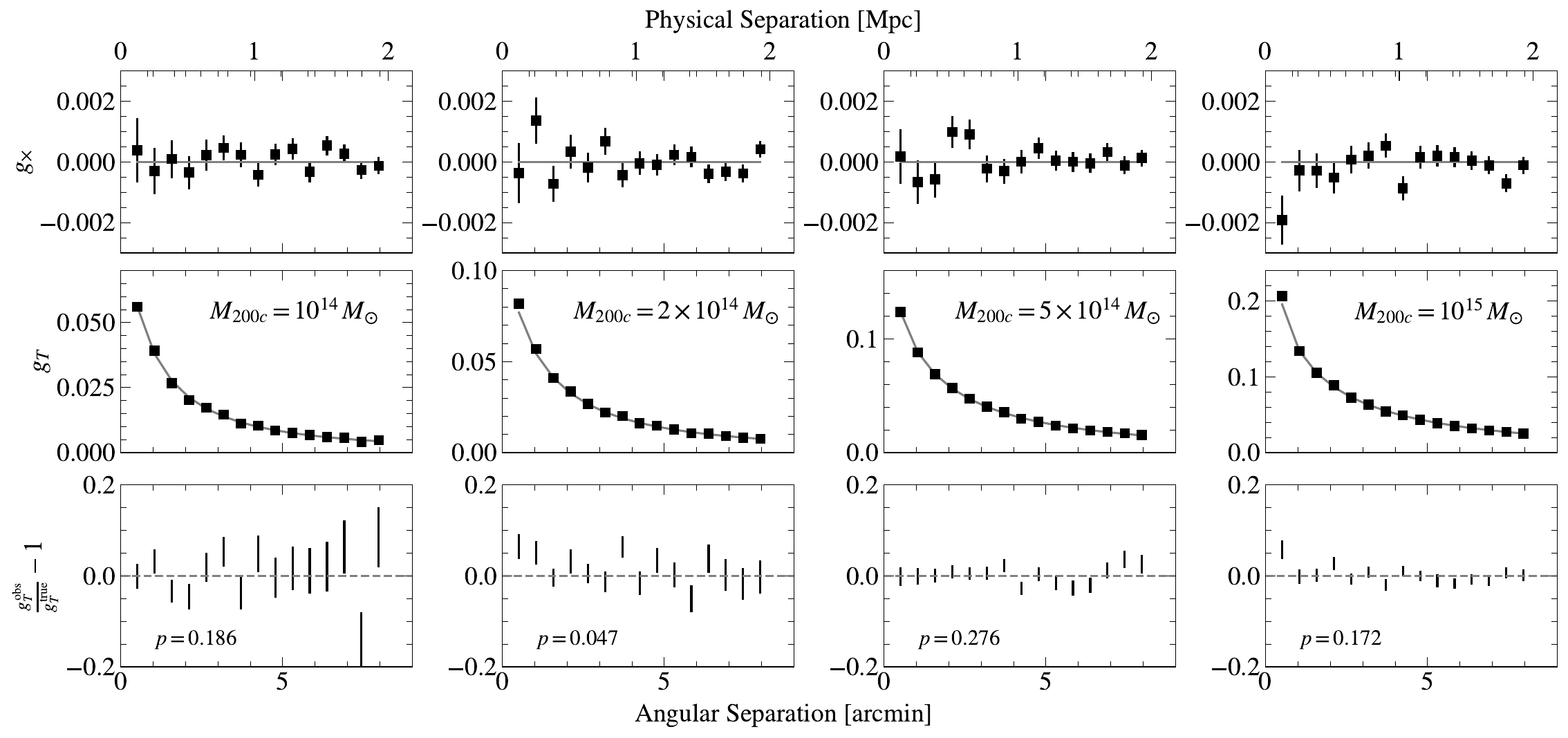}
    \caption{Shear estimation results for the cluster lens of masses $10^{14} \Msun$, $2 \times 10^{14} \Msun$, $5 \times 10^{14} \Msun$, and $10^{15} \Msun$. \anacal can recover input true shear up to $g \sim 0.15$. The error bar is from the forecast covariance for the 10-year LSST survey. We show the $p$-value in the last row of the figure. There is a positive shear bias that we will quantify in Section~\ref{sec:shear_bias_res}.}
    \label{fig:ex_shear_fit}
\end{figure*}

In Figure~\ref{fig:ex_shear_fit}, we show the shear estimation results for the cluster lens of masses $10^{14} \Msun, 2\times10^{14}\Msun, 5\times10^{14} \Msun$, and $10^{15} \Msun$ with the LSST 10-year covariance estimated in Section~\ref{sec:covariance}. In particular, we show the input cross and tangential shear and shear estimated by \anacal, and the ratio between the input shear and recovered shear. We find that \anacal can accurately recover the input tangential shear $g_T$ and cross shear $g_X$ across various cluster lens masses and radial ranges even in mildly non-linear regimes up to $g \sim 0.15$. This is consistent with the performance of \anacal in constant shear sims. On the very high shear end, we find a positive shear bias that is caused by blending that we will quantify in Section \ref{sec:shear_bias_res}.

\subsection{Shear bias}

\label{sec:shear_bias_res}
\begin{figure}[htbp]
    \includegraphics[width=\columnwidth]{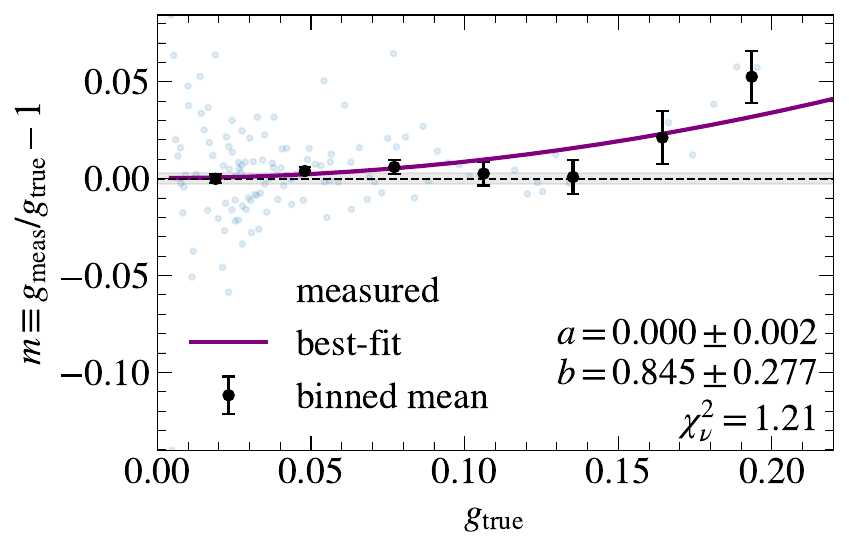}
    \caption{Measured and the best-fit shear bias for all radial bins of the simulated cluster lenses. The transparent points are measured shear bias, the solid circles with error bars are the binned mean of the measured shear bias, and the purple solid line is the best fit shear bias with Equation~\ref{eq:shear_bias}. The gray-shaded region is the SRD requirement for shear bias $|m|<0.3\%$. The best-fit parameters are printed in the lower right corner. The multiplicative bias is consistent with 0, and there is a positive third-order shear bias term.}
    \label{fig:shear_bias}
\end{figure}

In Figure~\ref{fig:shear_bias}, we show the fitted shear bias with the method described in Section~\ref{sec:shear_bias}. We find that the linear multiplicative shear bias is consistent with 0, while there is a significant third-order shear estimation bias with third-order bias coefficient $b=0.845 \pm 0.277$ . This is different from the results from constant shear simulations without blending \citep{liAnalyticalWeaklensingShear2023}, where the higher-order shear bias is negative. We find the same positive high-order shear bias in constant shear simulations with blending as shown in Appendix~\ref{appendix:shear_bias}. Therefore, we confirm that the positive high-order shear bias in our cluster lensing simulations is caused by blending, and the shear bias result in this paper is consistent with previous findings.

\subsection{Magnification effect on shear response}
\label{sec:response_trend}

\begin{figure}[htbp]
    \includegraphics[width=\columnwidth]{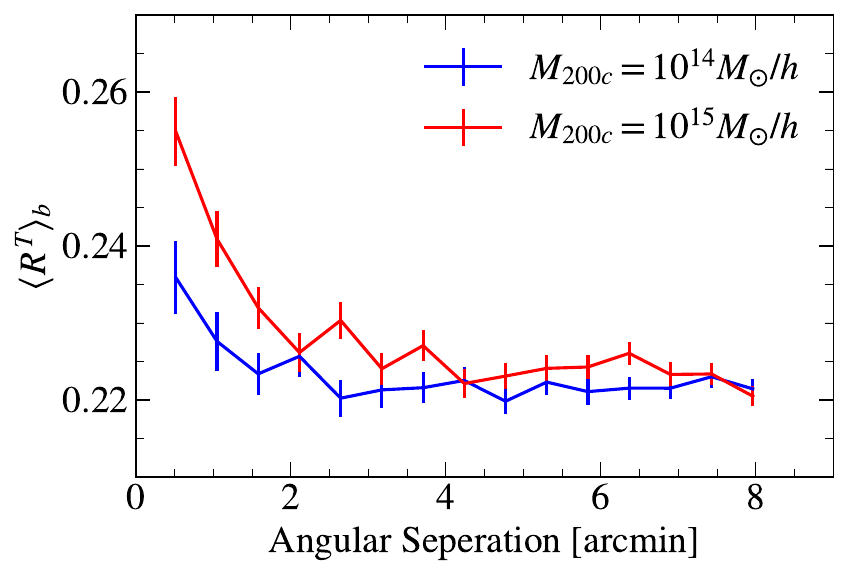}
    \caption{Source galaxy response as a function of the distance to cluster lens center. Near the cluster center, the mean response is significantly larger than in the cluster outskirts. We also observe a mass dependence of the response trend. We show that the mean response is correlated with mean convergence $\kappa$ in the angular bin in Figure~\ref{fig:response_trend}.}
    \label{fig:rt_angular}
\end{figure}

\begin{figure}[h]
    \includegraphics[width=\columnwidth]{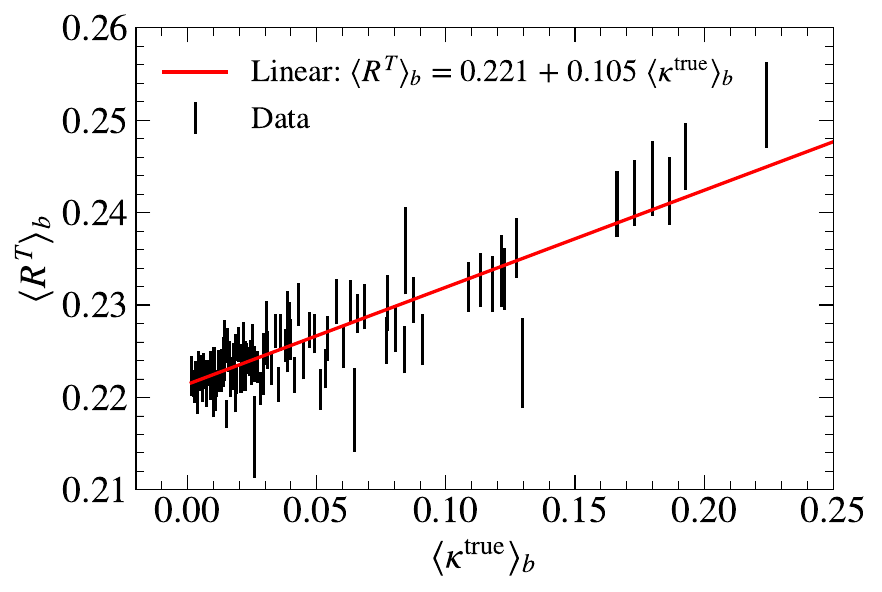}
    \caption{Mean \anacal tangential shear response in angular bins as a function of mean convergence. The mean shear response increases monotonically with the input convergence. We also test that when the simulations have no input $\kappa$, this trend goes away. Therefore, we think the shear response trend in \citet{grandisSRGEROSITAAllSky2024} is caused by the magnification effect. The red solid line is a linear fit of the mean response as a function of convergence. }
    \label{fig:response_trend}
\end{figure}

In \citet{grandisSRGEROSITAAllSky2024}, the authors reported a radial variation in the \mcal shear response that shows a mild dependence on richness. We find the same qualitative behavior in our cluster simulations. Figure~\ref{fig:rt_angular} shows the mean \anacal shear response measured in annular bins for cluster lenses with masses $10^{14}\,\Msun$ and $10^{15}\,\Msun$. In both cases, the response rises markedly toward the cluster center.

We verify that this trend is driven by lensing magnification. Background sources are magnified by
\begin{equation}
\mu=\frac{1}{(1-\kappa)^2-|\gamma|^2},
\end{equation}
which increases their apparent sizes. Because the shear response depends on source size, a non-negligible convergence field ($\kappa$) boosts the measured response through magnification. Figure~\ref{fig:response_trend} compares the mean response in annular bins to the mean input convergence in the same bins, showing a clear monotonic increase of response with $\kappa$. For the cluster lens with mass $\mtwoh = 10^{15} \Msun$, the average response at 0.5 arcmin is $\sim 15\%$ higher than the average response in the outskirts. We also see there is a significant trend with halo mass and, by extension, richness. As a further check, when we repeat the simulations with the convergence field removed, the radial response trend disappears.

This radial dependence has an important practical implication for cluster shear inference: using a single, globally averaged response for all sources would miscalibrate the shear profile—overestimating shear at small radii (where the true response is higher) and underestimating it at larger radii. In Section~\ref{sec:mass_bias}, we quantify the resulting bias in cluster mass estimates when this magnification-driven response variation is neglected.

\subsection{Mass bias}
\label{sec:mass_bias}
\begin{figure}[h]
    \includegraphics[width=\columnwidth]{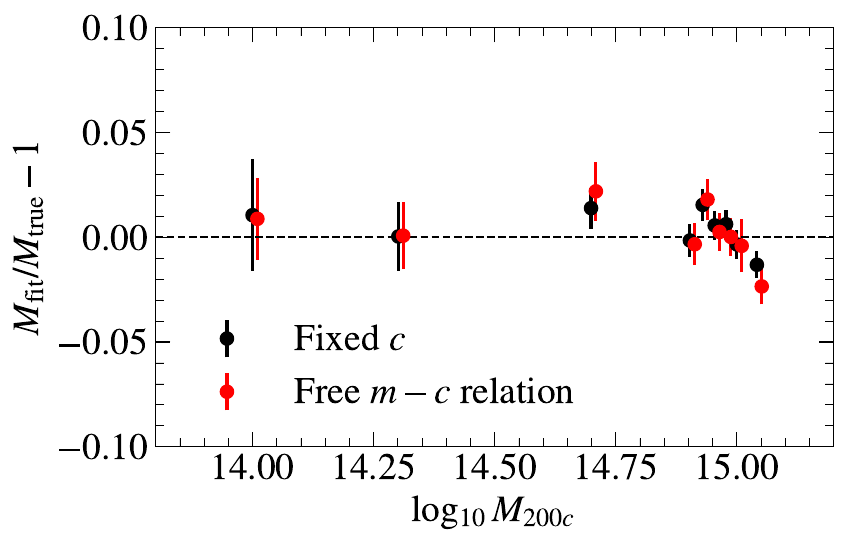}
    \caption{Best fit mass compared to the true input mass for all simulated cluster lenses with a scale cut of 0.2 Mpc at $z=0.25$. The black points use true concentration in mass fitting, while the red points fit mass and concentration simultaneously. The standard deviation of mass estimation bias is 0.8\% for the fixed true concentration case and 1.2\% for the free mass-concentration relation case.  The error bars in this plot stand for the uncertainty of the mass estimation bias based on our finite simulations. We correct for the radial response trends in this plot.
    }
    \label{fig:mass_bias}
\end{figure}

\begin{figure}[h]
    \centering
    \includegraphics[width=\columnwidth]{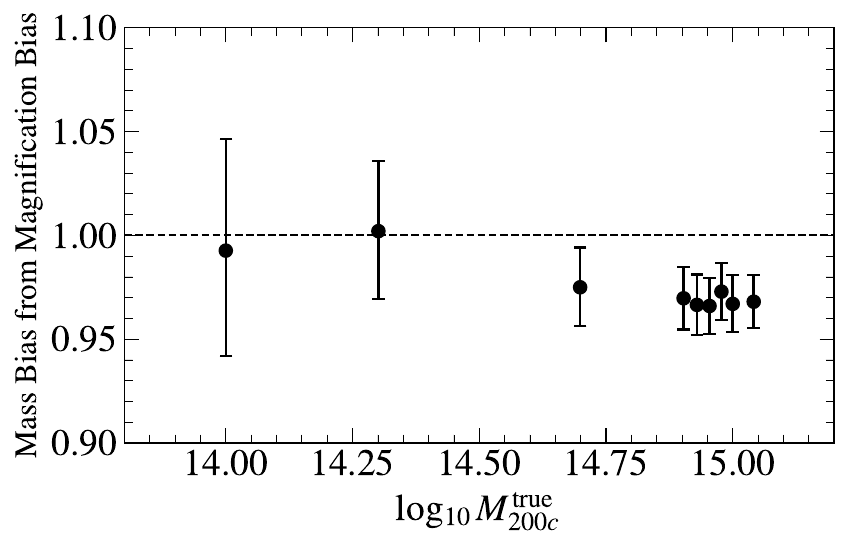}
    \caption{Mass bias caused by response trend shown in Section~\ref{sec:response_trend}. The y-axis is the ratio between the best-fit mass using the mean response of all radial bins and the best-fit mass using the mean response in each radial bin using true concentration. We find that on average, the radial response trend would cause a $\sim 3\%$ bias on mass calibration of high mass clusters if not treated correctly.}
    \label{fig:response_mass_bias}
\end{figure}

\begin{figure}[h]
    \includegraphics[width=\columnwidth]{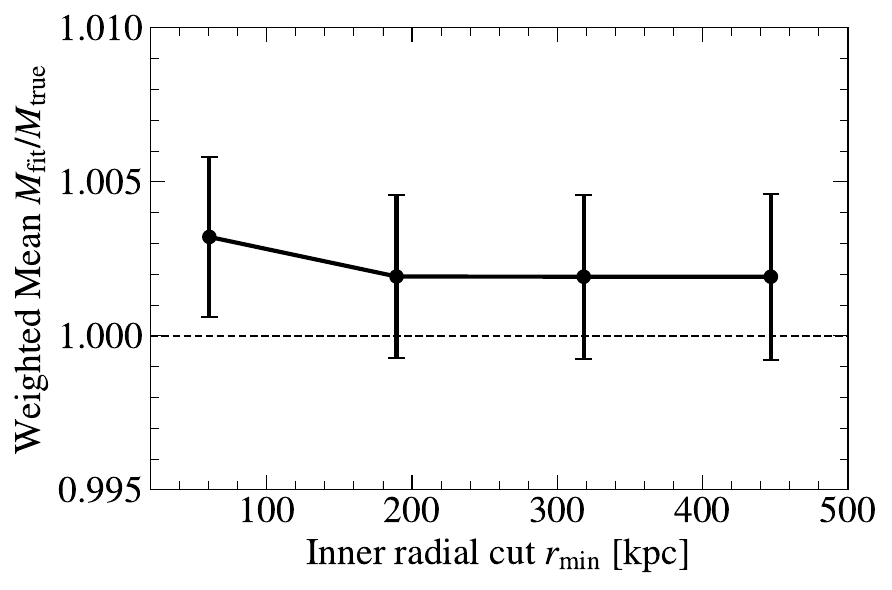}
    \caption{Mean mass calibration bias across all mass range as a function of radial scale cut. Each mass calibration result is weighted by its bootstrap uncertainty, and the error bar on the mean is propagated accordingly. The mass calibration result when including lensing below 0.18 Mpc at $z=0.25$ is biased high because of the positive shear bias. We see that at scales above 0.2 Mpc at $z=0.25$, \anacal's mean cluster mass calibration bias is smaller than 0.25\%.}
    \label{fig:radial_cut}
\end{figure}

In Figure~\ref{fig:mass_bias}, we show the ratio between the best-fit masses from \anacal shear estimation and true input masses. For each cluster mass, we bootstrap resample cluster simulations to estimate the variance and mean mass profile because of the finite number of simulations. The statistical uncertainty for each mass is estimated by fitting a mass to the bootstrapped mean shear profile with the analytic covariance matrix representing realistic lensing covariance in data, and thus the error bar in Figure~\ref{fig:mass_bias} stands for the uncertainty of our estimation of the mass estimation bias. The standard deviation of the estimated mass bias corresponds to the uncertainty that we should expect for stacked cluster lensing in any richness bin. Having more image simulations would shrink the error bar for the estimation of mass bias but would not change the bias itself. When fixing the input concentration to the true concentration, across all mass bins, the standard deviation of mass estimation bias is 0.8\%. When relaxing the mass and concentration relation, across all mass bins, the standard deviation of the mass estimation bias is 1.2\%. In Figure~\ref{fig:response_mass_bias}, we show the impact of the radial response trend described in Section~\ref{sec:response_trend} on mass calibration. We see that across all mass ranges, not taking into account the radial response trend additionally adds a $\sim 3\%$ mass calibration bias.

Cutting the regions near the cluster center can help avoid the high-shear and high-convergence region \citep{okabeLoCuSSWeaklensingMass2016, mcclintockDarkEnergySurvey2019}. In Figure~\ref{fig:radial_cut}, we show the averaged mass estimation bias across all mass ranges as a function of radial cut. Each mass is weighted by its mass bias uncertainty estimated with bootstrap resampling, and the uncertainty on the mean bias is propagated accordingly. We find that the mass calibration result when including lensing within 0.2 Mpc is slightly biased high because of the positive shear bias we discussed in Section~\ref{sec:shear_bias_res}. We see that at scale above 0.18 Mpc at $z=0.25$, \anacal's mean cluster mass calibration bias is smaller than 0.25\%, since we excluded source galaxies in the high shear regime and the shear estimation near the cluster center is down-weighted by the covariance matrix.

\section{Summary}
\label{sec:summary}
In this work, we simulated cluster lenses in the setting of the 10-year LSST observations. Our key findings are

\begin{itemize}

    \item The magnification effect causes a non-negligible radial response trend. If we use the average response across all radial bins, the shear estimation of inner radial bins will be biased low while the outer bins will be biased high, and the mass and concentration will also be biased.

    \item From low shear to mildly high shear $g \in (0, 0.15]$, \anacal can accurately recover the input shear well. This shear range covers most of the cluster lens shear amplitude for stacked cluster lensing analysis in a reasonable radial range. We do not simulate high-order distortions like flexions in this work. However, in the radial range dealt with in this paper, the amplitude of flexion is small ($\leq 0.01$), and it is unlikely to change the conclusion. It is possible that substructures in clusters can cause a strong flexion field. In the future, we plan to use ray-tracing simulations to study the effect of high-order distortions \citep{liuMeasurementCalibrationNonlinear2024}.

    \item The high shear values near cluster centers induce a positive third-order shear bias. We further confirm this behavior using constant-shear blended simulations, obtaining consistent results.

    \item Under the ideal situation with the source and lens redshifts and mass-concentration relation perfectly known, no blending between cluster members and source galaxies or between source galaxies of different redshifts, the scatter of mass recovery bias is about 1\% for cluster lenses weighing $10^{14} \Msun$ to $10^{15} \Msun$.

    \item Restricting the analysis to $R \ge 0.2\,\mathrm{Mpc}$, the mean mass bias from \anacal\ is below $1\%$: the strongly biased inner region is excluded, and any residual contribution from the high-shear core is further suppressed by covariance weighting. In contrast, including shear measurements at $R < 0.2\,\mathrm{Mpc}$ leads to a slightly high bias in the inferred mass, driven by the positive shear bias near the cluster center.
\end{itemize}

Because of the isolation of high shear from other complexities in this work, we gain an understanding of the bias of the optimal cluster mass calibration only from high shear and in the absence of other systematics. This work forms a baseline for the cluster mass calibration for LSST. Moreover, our assessment can also be applied to methods that directly forward model lensing instead of doing mass calibration \citep{parkClusterCosmologyAnisotropic2023,salcedoDarkEnergySurvey2023, zhouForecastingConstraintsOptical2024}.

In the future, we plan to incorporate more realistic conditions, including cluster member galaxies, intra-cluster light, source redshift distribution, cluster substructure, high order distortions, more realistic mass and concentration distributions, and expand the error budget we obtain in this work. With the careful and comprehensive studies and tests of the resulting bias and variance of various modeling choices and systematics for cluster lensing, cluster cosmology will be an exciting precision science venue for Stage IV surveys.

The methodology presented here, which estimates systematic biases via realistic image simulations, extends beyond cluster mass estimation. This simulation infrastructure can be leveraged to validate flux recovery, redshift estimation, and measurements for galaxy clustering and cosmic shear. Ultimately, this method would enable a comprehensive pixel-to-cosmology multi-probe test of the analysis pipeline by comparing input parameters with inferred values. This represents the most stringent validation for survey cosmology, capturing systematics from the CCD level down to cosmology parameter inference.

\begin{acknowledgments}
This paper has undergone internal review by the LSST
Dark Energy Science Collaboration. The internal review-
ers were Anthony Englert and Miranda Gorsuch.

Xiangchong Li acknowledges support from the U.S. Department of Energy under
Contract No. DE-SC0012704 and from the Laboratory Directed Research and
Development (LDRD) Program at Brookhaven National Laboratory (Project No.
27992). We thank Doug Clowe for useful conversations.
The DESC acknowledges ongoing support from the Institut National de
Physique Nucl\'eaire et de Physique des Particules in France; the
Science \& Technology Facilities Council in the United Kingdom; and the
Department of Energy and the LSST Discovery Alliance
in the United States.  DESC uses resources of the IN2P3
Computing Center (CC-IN2P3--Lyon/Villeurbanne - France) funded by the
Centre National de la Recherche Scientifique; the National Energy
Research Scientific Computing Center, a DOE Office of Science User
Facility supported by the Office of Science of the U.S.\ Department of
Energy under Contract No.\ DE-AC02-05CH11231; STFC DiRAC HPC Facilities,
funded by UK BEIS National E-infrastructure capital grants; and the UK
particle physics grid, supported by the GridPP Collaboration.  This
work was performed in part under DOE Contract DE-AC02-76SF00515.

\end{acknowledgments}

\begin{contribution}
The contributions of the primary authors are as follows. CZ and XL developed the simulation. CZ analyzed the simulations and wrote the paper. HW calculated the covariance. AvdL, TJ, TS, SB, TS, SF, PA, LB, and SM reviewed the paper and provided comments. AE and MG served as internal reviewers of the paper. AAPM contributed to the software that was central to the paper.
\end{contribution}

\appendix

\section{Shear bias in constant shear simulations}
\label{appendix:shear_bias}

\begin{figure*}[h]
    \centering
    \includegraphics[width=0.7\textwidth]{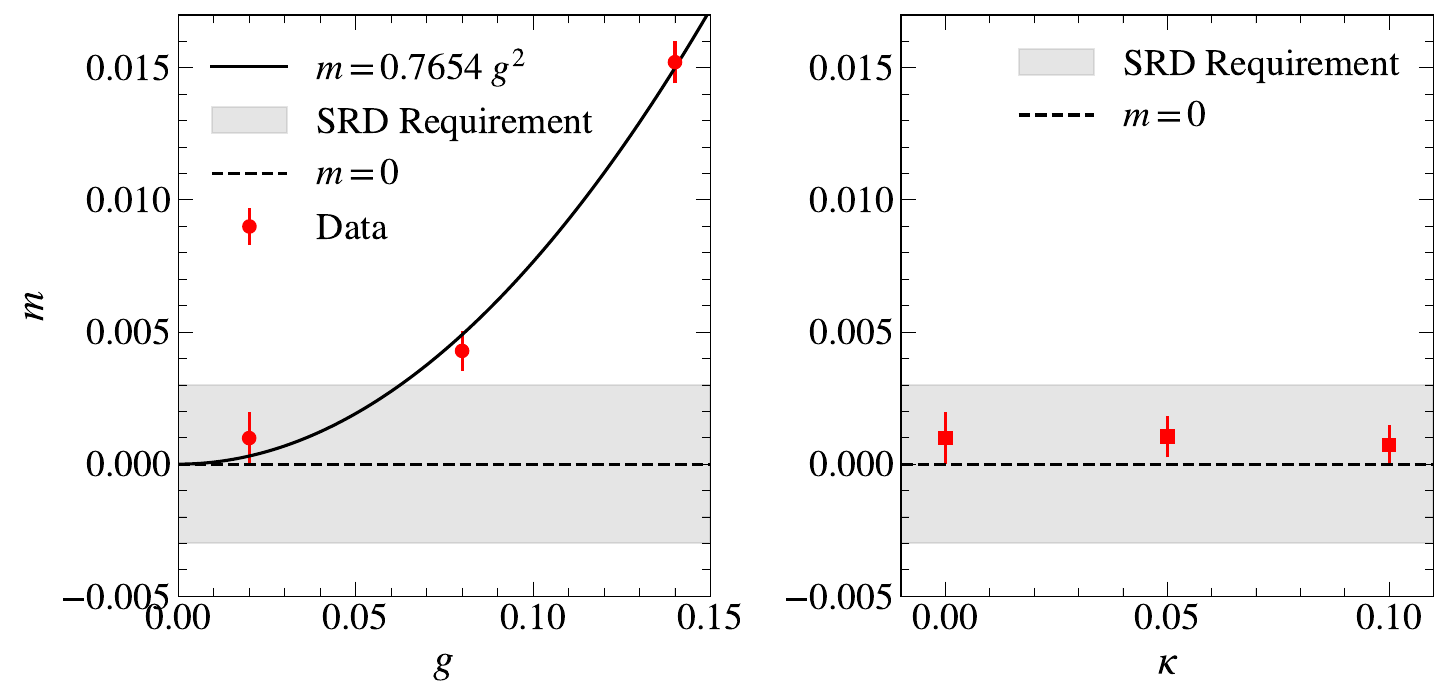}
    \caption{The high-order shear bias in constant blended simulations. {\bf Left: } Multiplicative bias as a function of input shear. There is a positive third-order shear bias.  {\bf Right:} Multiplicative bias as a function of input convergence. We find that convergence does not cause a shear bias. This is consistent with the cluster lens result discussed in Section~\ref{sec:shear_bias}.}
    \label{fig:constant_shear_bias}
\end{figure*}

In Figure~\ref{fig:constant_shear_bias}, we present the higher-order shear biases measured from constant-shear simulations with blended source galaxies. We simulate LSST-like $i$-band images of $4000 \times 4000$ pixels at
$0\farcs2$ pixel$^{-1}$, populated with a blended source galaxy population.  A uniform reduced shear
$g$ is applied across each image using three sign modes
($g = +|g|$, $g = -|g|$, and $g = 0$) so that shape-noise contributions
cancel in the difference estimator.  We repeat this procedure at several
input shear amplitudes ($g = 0.02$, $0.08$, and $0.14$) and at several
convergence values ($\kappa = 0.00$, $0.05$, and $0.10$) to separately
disentangle the dependence of the multiplicative bias on shear amplitude
and convergence. We find a significant contribution at the third order in the input shear. In addition, the convergence shows no correlation with the linear-order shear bias, consistent with the prediction of \citet{zhangMeasuringReducedShear2011}.

\software{
    astropy \citep{2013A&A...558A..33A,2018AJ....156..123A,2022ApJ...935..167A},
    colossus \citep{diemerCOLOSSUSPythonToolkit2018}, numpy
    \citep{harrisArrayProgrammingNumPy2020}, scipy
    \citep{virtanenSciPy10Fundamental2020}, matplotlib
    \citep{hunterMatplotlib2DGraphics2007},
}

\bibliography{zotero, reference}{}
\bibliographystyle{aasjournalv7}

\end{document}